# Non-Hermitian $C_{NH}$ = 2 Chern insulator protected by generalized rotational symmetry

Kai Chen[1,2,3,©], Alexander B. Khanikaev[1,2,3,*]

[1]Department of Physics, City College of New York, New York, NY 10031, USA.
[2]Department of Electrical Engineering, Grove School of Engineering, City College of the City University of New York, 140th Street and Convent Avenue, New York, NY 10031, USA.
[3]Physics Program, Graduate Center of the City University of New York, New York, NY 10016, USA.

We propose a non-Hermitian topological system protected by the generalized rotational symmetry which invokes rotation in space and Hermitian conjugation. The system, described by the tight-binding model with reciprocal imaginary next-nearest-neighbor hopping, is found to host two pairs of in-gap edge modes in the gapped topological phase, and is characterized by the non-Hermitian (NH) Chern number $C_{NH} = 2$. The quantization of the non-Hermitian Chern number is shown to be protected by the generalized rotational symmetry $\hat{H}^+(g\vec{k}) = \hat{U}_g \hat{H}(\vec{k}) \hat{U}_g^+$ of the system. Our finding paves the way towards novel non-Hermitian topological systems characterized by large values of topological invariants and hosting multiple in-gap edge states, which can be used for topologically resilient multiplexing.

**Introduction**

Topological phases represent a new state of matter which extends beyond the conventional symmetry-based Landau paradigm[1,2]. The first, and perhaps the most recognized, topological phase in modern physics is the integer quantum Hall insulator (IQHI)[3,4]. IQHI is characterized by an integer number of edge states at the interface between the quantum Hall insulator and a trivial insulator, with the number of edge states and the quantized edge conductance determined by the Chern number, the topological invariant which is totally defined by the bulk band topology of the system[5]. Since the discovery of the IQHI, the topological concepts started to play an increasingly important role in our understanding and classification of materials [6,7], blossoming into a new research field of condensed matter physics, and even spilling into classical domains of topological acoustics, mechanics, and photonics[8-11].

The celebrated Altland-Zirnbauer (AZ) "10-fold way" [12] allows to discern any topological quantum system by its pattern of nonspatial symmetries, i.e., Chiral symmetry, particle-hole symmetry and time reversal symmetry. Taking into consideration spatial symmetries allowed introducing the concept of topological crystalline phases[13] characterized by gapped ground states that are not adiabatically connected to an atomic limit as long as certain spatial symmetries, which include mirror and rotational symmetries, are preserved. While originally limited to purely Hermitian systems, in the recent years, the non-Hermitian models has brought new flavors to topological physics [14-23]. The non-Hermitian systems were found to host novel phenomena, including without direct Hermitian analogues. One example is the generalized Bulk-edge correspondence[24-28], which represents a guiding principle of Hermitian topological matter, and can break down in the non-Hermitian settings. For instance, a system with open boundary conditions cannot be understood by studying the band topology of the system with periodic boundary condition[29]. Other examples include new class of "skin" boundary states[30-32] and non-Hermitian exceptional points[33-35] unique to non-Hermitian systems, and even totally new topological phases enabled by the additional imaginary dimension of the energy spectrum [36,37]. The 10-fold AZ symmetry classification in Hermitian systems has been recently extended to a 38-fold classification in non-Hermitian systems to encompass the distinction between complex conjugation and transposition for non-Hermitian Hamiltonian [23]. We note, however, that the generalized rotational symmetry proposed here, $\hat{H}^+(g\vec{k}) = \hat{U}_g \hat{H}(\vec{k})\hat{U}_g^+$, does not belong to this extended 38-fold symmetry classification. This is dues to the fact that the respective 38-fold symmetry classification only considers non-spatial symmetries. Introduction of additional spatial symmetries would require a further expansion of the topological classification[22], a direction worth of a separate rigorous study.

While some non-Hermitian generalizations of the IQHI have been reported in literature[38,39], the systems considered so far were characterized by a non-Hermitian Chern number $C_{NH} = \pm 1$ and thus could only allow existence of a single edge state (or a pair of edge states for the case of domain wall geometry) in the topologically non-trivial gap. On the other hand, higher values of topological invariants can be of interest and may also be highly desirable for numerous practical applications due to the possibility of multiplexing via several boundary modes[40]. Thus, it is worth exploring the new physics and the possibility of a non-Hermitian Chern insulator with higher non-Hermitian Chern number $|C_{NH}| \geq 2$, which, to the best of our knowledge, have not yet been discovered.

In this work, we build upon prior work on Hermitian model[41] characterized by higher Chern number values, and we propose a non-Hermitian system with higher non-Hermitian topological invariant, the Chern number $C_{NH} = 2$. The system is described by a non-Hermitian tight-binding model which is shown to host two unidirectional edge states at a single boundary within a topological gap. We prove that the reported topological phase and the edge states are protected by a generalized rotational symmetry which invokes both spatial rotation and Hermitian conjugation. Specifically, we find that the generalized rotation symmetry for our proposed non-Hermitian Hamiltonian ensures quantization of the non-Hermitian Chern number, which, in turn, defines the net number of edge modes, thus establishing a generalized non-Hermitian bulk boundary correspondence. Similar to the case of Hermitian Chern insulators, the non-Hermitian Chern number does not allow predicting the boundary behavior of the finite size model for the case of gapless bulk bands, because the Berry connection is not well defined.

**The tight-binding model and an elevated non-Hermitian Chern number**
In what follows, we consider a non-Hermitian tight-binding model with complex next-nearest-neighbor (NNN) coupling and depicted in Fig. 1a. This model is found to exhibit two distinct gapped phases, topological and trivial gapped phases, which are separated by the gapless phase. The topological gapped phase hosts two pairs of in-gap edge modes. As detailed below, the topological phase and the edge states are protected by the generalized rotational symmetry for the non-Hermitian Hamiltonian $\hat{U}_g \hat{H}(g\vec{k})\hat{U}_g^+ = \hat{H}^+(\vec{k})$, where $\hat{U}_g$ is the unitary representation of the generator $g \equiv R_z(\frac{\pi}{2})$ of the four-fold rotational group $C_4$. In our model it is the unitary matrix $\hat{U}_g \equiv e^{-\frac{i\pi}{2}\hat{\sigma}_z}$.

The generalized rotational symmetry quantizes the non-Hermitian Chern number (see Supplemental Material for details), which assumes value $C_{NH} = 2$ in topological gapped phase, which explains the emergence of a pairs of in-gap edge states at each boundary of the system. The non-Hermitian Chern number used here represents a direct analogue of the Hermitian Chern number and can be expressed in the form $C_{NH} \equiv \frac{1}{2\pi i}\iint_{BZ} \vec{B}_n(\vec{k})\, dk^2$, where $\vec{B}_n(\vec{k}) = \nabla_{\vec{k}} \times \vec{A}_n(\vec{k})$ is the non-Hermitian Berry curvature. The difference with the Hermitian version lays in the fact that the non-Hermitian Berry connection should be evaluated with the use of the left and right eigenstates of the non-Hermitian Hamiltonian, i.e., $\vec{A}_n(\vec{k}) \equiv <\phi_n(\vec{k})|\nabla_{\vec{k}}|\psi_n(\vec{k})>$, where the vector $<\phi_n(\vec{k})|$ ($|\psi_n(\vec{k})>$) is the left (right) eigenstate of the non-Hermitian Hamiltonian $\hat{H}(\vec{k})$, i.e., $<\phi_n(\vec{k})|\hat{H}(\vec{k}) = \epsilon_n(\vec{k}) <\phi_n(\vec{k})|$ and $\hat{H}(\vec{k})|\psi_n(\vec{k})> = \epsilon_n(\vec{k})|\psi_n(\vec{k})>$, and where $n$ is the band index. We would like to emphasize one important property of the non-Hermitian matrix, that its left (right) eigenstates corresponding to different eigenvalues may not be orthogonal with each other, however, the left eigenstate with eigenvalue $\epsilon_1$ and the right eigenstate with eigenvalue $\epsilon_2$ are orthogonal as long as $\epsilon_1 \neq \epsilon_2$.

The tight-binding Hamiltonian of the proposed system (Fig. 1(a)) in real space is defined as

$$\hat{H} = m_0 \sum_{m,n}(a^+_{m,n} a_{m,n} - b^+_{m,n} b_{m,n})$$
$$+ t_x \sum_{m,n}[(a^+_{m,n} b_{m+1,n} + a^+_{m+1,n} b_{m,n} - a^+_{m,n} b_{m,n+1} - a^+_{m,n+1} b_{m,n}) + h.c.]$$
$$+ t_y \sum_{m,n}[\frac{i}{2}(b^+_{m,n} a_{m+1,n-1} + b^+_{m,n} a_{m-1,n+1} - b^+_{m,n} a_{m+1,n+1} - b^+_{m,n} a_{m-1,n-1}) + h.c.]$$
$$+ t_z \sum_{m,n}[(a^+_{m,n} a_{m+1,n} + a^+_{m,n} a_{m,n+1} - b^+_{m,n} b_{m+1,n} - b^+_{m,n} b_{m,n+1}) + h.c.]$$
$$- \frac{i\gamma_z}{2} \sum_{m,n}(a^+_{m,n} a_{m+1,n+1} + a^+_{m,n} a_{m-1,n-1} - a^+_{m,n} a_{m+1,n-1} - a^+_{m,n} a_{m-1,n+1}$$
$$+ b^+_{m,n} b_{m+1,n-1} + b^+_{m,n} b_{m-1,n+1} - b^+_{m,n} b_{m+1,n+1} - b^+_{m,n} b_{m-1,n-1}). \quad (1)$$

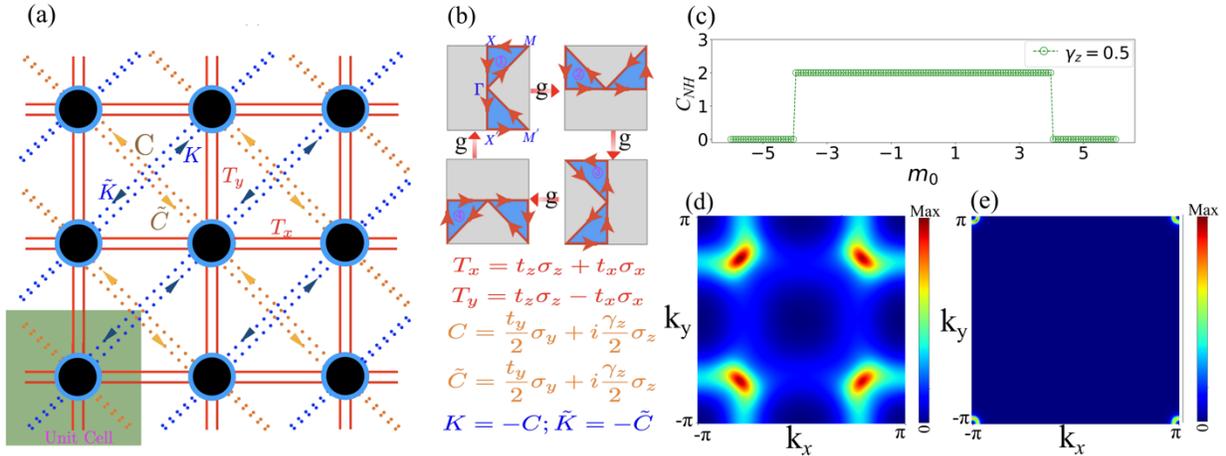

**FIG. 1.** (a) Schematic of the proposed non-Hermitian tight-binding model. (b) The first Brillouin zone (BZ) and high symmetric points of the model. (c) The non-Hermitian Chern number $C_{NH}$ as a function of parameter $m_0$ with $\gamma_z = 0.5$, and $t_x = t_y = t_z = 1$. (d, e) Imaginary part of the Berry curvature distribution over the first BZ for the occupied band for (d) $m_0 = 1$ and (e) $m_0 = 3.95$, close to topological transition.

In the tight-binding model described by Eq (1) we assumed that there are two degrees of freedom (represented by states $A$ and $B$) at each site, with $a_{m,n}$ ($b_{m,n}$) and $a^+_{m,n}$ ($b^+_{m,n}$) representing annihilation and creation operators for the state $A$ ($B$). States $A$ and $B$ can be of different energy $m_0$ and $-m_0$, and $t_x$ is the nearest neighbor (NN) hopping amplitude between the states $A$ (states

$B$) of the neighboring unit cells, $t_y$ is the NNN hopping amplitude between state $A$ and state $B$, $t_z$ is the NN hopping amplitude between state $A$ ($B$) and itself. The $-i\gamma_z$ ($i\gamma_z$) represent complex NNN hopping between $A$ states (and $B$ states), which introduces non-Hermicity into our tight-binding model. In this paper, we set $t_x = t_y = t_z = 1$ without loss of generality. The Hamiltonian Eq. (1) can be expressed in the momentum space as

$$\hat{H}(k_x, k_y) = 2t_x(\cos k_x - \cos k_y)\hat{\sigma}_x + t_y[\cos(k_x - k_y) - \cos(k_x + k_y)]\hat{\sigma}_y$$
$$+ [m_0 + 2t_z(\cos k_x + \cos k_y) + i2\gamma_z \sin k_x \sin k_y]\hat{\sigma}_z, \qquad (2)$$

where $\{\hat{\sigma}_x, \hat{\sigma}_y, \hat{\sigma}_z\}$ are Pauli matrices acting in the Hilbert space spanned by the states A and B, and the momentum $\vec{k} = (k_x, k_y)$ is mapped to $(gk_x, gk_y) \equiv R_z\left(\frac{\pi}{2}\right)(k_x, k_y) = (k_y, -k_x)$ under counter-clockwise $\frac{\pi}{2}$ rotation around $z$ axis. It's not hard to see for the Hamiltonian in the momentum space that it possesses the generalized rotational symmetry defined as $\hat{H}^+(gk_x, gk_y) = \hat{U}_g \hat{H}(k_x, k_y) \hat{U}_g^+$ with $\hat{U}_g = e^{-\frac{i\pi}{2}\hat{\sigma}_z}$, and one can also show that the Hamiltonian with opposite momentum $\vec{k}$ can be obtained by applying the operator twice, e.g., $\hat{U}_g^2$.

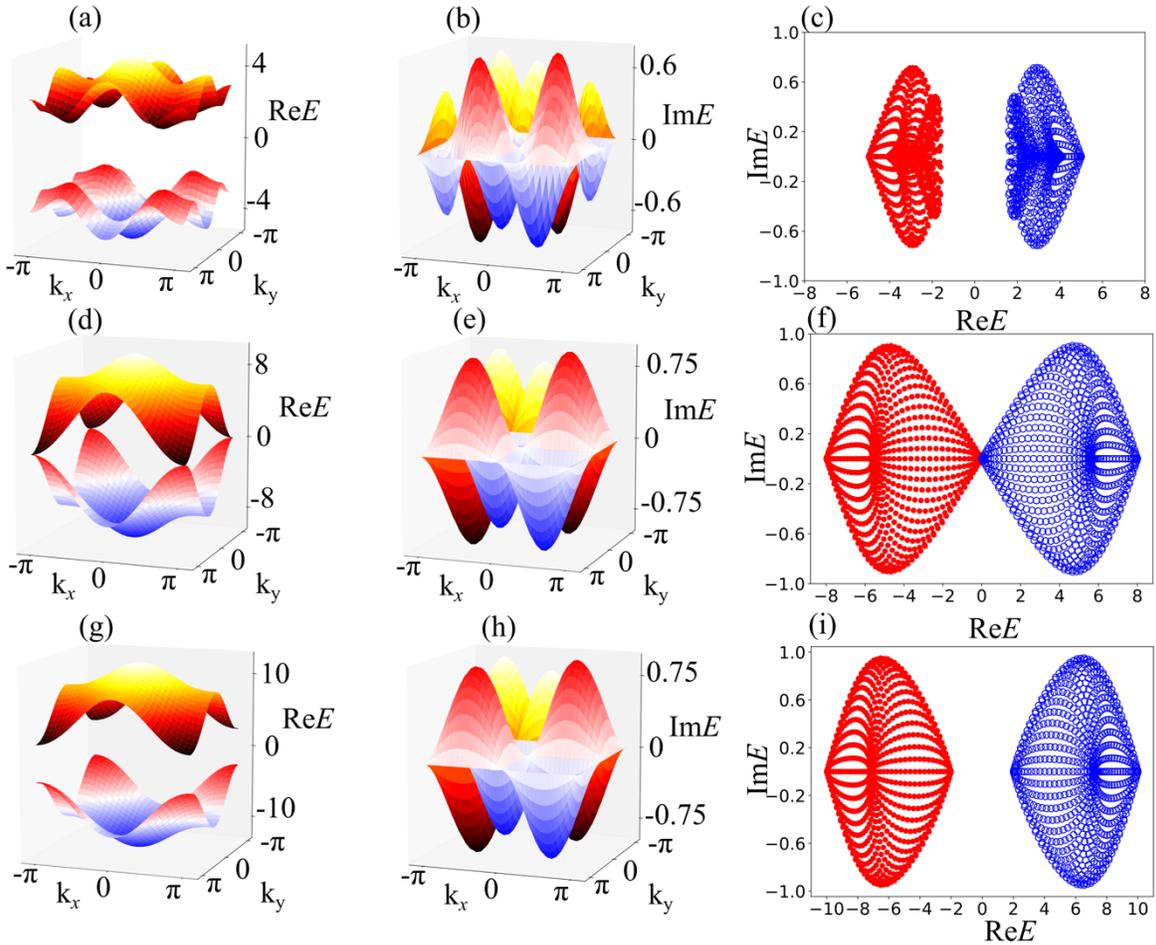

**FIG. 2.** (a-b) Real and imaginary parts of the bulk energy spectra for $m_0 = 1$ (topological phase), and (c) respective complex energy spectra of two bulk bands (red and blue regions) in the complex-energy plane for $m_0 = 1$. (d-e) Real and imaginary parts of the bulk energy spectra for $m_0 = 4$ (gapless phase) and (f) respective complex energy spectra of two bulk bands for $m_0 = 4$. (g-h) Real and imaginary parts of bulk energy spectra for $m_0 = 6$ (trivial phase), and (i) respective complex energy spectra of two bulk bands for $m_0 = 6$.

We would like to emphasize that the systems described by the Hamiltonian Eqs. (1) and (2) can be brought to gapless phase by properly tuning the parameters $\gamma_z$ and $m_0$, however, in this work, we are interested in the gapped phase of non-Hermitian Chern insulator with higher non-Hermitian Chern number and multiple edge modes, and the case of the Chern number $C_{NH} = 2$ specifically. Therefore, we focus on the values of parameters that ensure the existence of a line gap in the complex energy plane.

In our model, due to the form of the symmetry operator $\hat{U}_g$, the non-Hermitian Chern number can be calculated by evaluating the expectation value of $\hat{\sigma}_z$ with respect to the right eigenstates at the high symmetry points in the first Brillouin zone:

$$C_{NH} = \frac{2}{\pi} \, Im \oint_\gamma \vec{A}_n(\vec{k}) d\vec{k} = \frac{2}{\pi} \left[ 2\theta_g(X) - 3\theta_g(\Gamma) + \theta_g(M) \right] \; (\text{mod } 4), \tag{3}$$

where $\gamma$ represent the loop in the first Brillouin zone, as shown in Fig.1(b), $\theta_g(\vec{K}) = \frac{\pi}{2} < \psi_n(g\vec{K})|\hat{\sigma}_z|\psi_n(\vec{K})>$, $\vec{K} \in \{\Gamma, X, M\}$, and $|\psi_n(\vec{K})>$ is the right eigenstate corresponding to eigenvalue with the negative real part.

We first calculate the non-Hermitian Chern number as function of $m_0$ with $\gamma_z = 0.5$. As shown in Fig. 1(c), for $m_0 \in (-4, 4)$, the system is in the topological gapped phase with the non-Hermitian Chern number $C_{NH} = 2$, and for other values of $m_0$, the non-Hermitian Chern number vanishes $C_{NH} = 0$, bringing the system to the trivial phase. At $m_0 = \pm 4$, the gap closes and the system transitions from the topological gapped phase to the trivial gapped phase, as shown in Fig. 2.

The generalized rotation symmetry of our non-Hermitian model restricts the Berry connection in the first Brillion zone to satisfy the following equation

$$\vec{A}_n(g^l\vec{k}) = \begin{cases} g^l[\vec{A}_n(\vec{k}) + i\, l\, \nabla_{\vec{k}}\, \theta_g(\vec{k})] & for\; l \in 2\mathbb{Z} \\ g^l[-\vec{A}_n^*(\vec{k}) + i\, l\, \nabla_{\vec{k}}\, \theta_g(\vec{k})] & for\; l \in 2\mathbb{Z}+1 \end{cases}. \tag{Eq.4}$$

For a non-Hermitian system, the respective non-Hermitian Berry curvature $\vec{B}_n(\vec{k}) = \nabla_{\vec{k}} \times \vec{A}_n(\vec{k})$ assumes complex values in general, therefore, the non-Hermitian Chern number $C_n \equiv \frac{1}{2\pi i} \iint_{BZ} \vec{B}_n(\vec{k}) \, d\vec{k}^2$ is complex. However, according to Eq.4, the real part of the non-Hermitian Berry curvature at $\vec{k}$ cancel with the real part of the non-Hermitian Berry curvature at $R_z(\frac{\pi}{2})\vec{k}$, leading to the net zero value of the imaginary part of the non-Hermitian Chern number.

Inspection of the imaginary part of the non-Hermitian Berry curvature in the BZ for the topological gapped phase reveals four peaks, shown in Fig. 1(d,e), with each peak contributing $\frac{1}{2}$ to the real part of the non-Hermitian Chern number. At the same time, Eq.4 ensures that there exist four equivalent peaks in the imaginary part of the non-Hermitian Berry curvature, as shown in Fig. 1(d,e), thus yielding total non-Hermitian Chern number of 2 in the topological gapped phase. Fig.1(d) and Fig.1(e) reveal how the position of the imaginary part of the non-Hermitian Berry curvature changes with the parameter $m_0$, indicating that there always exist four equivalent peaks in the first Brillion zone for the topological gapped phase. This evolution clearly shows that the higher value of the Chern number originates from the four corners of the Brillion zone, which are not equivalent due to the non-Hermitian character of the system, thus allowing these corners to manifest as four effective valleys (regions in the momentum space) each contributing equal value of ½ to the net non-Hermitian Chern number.

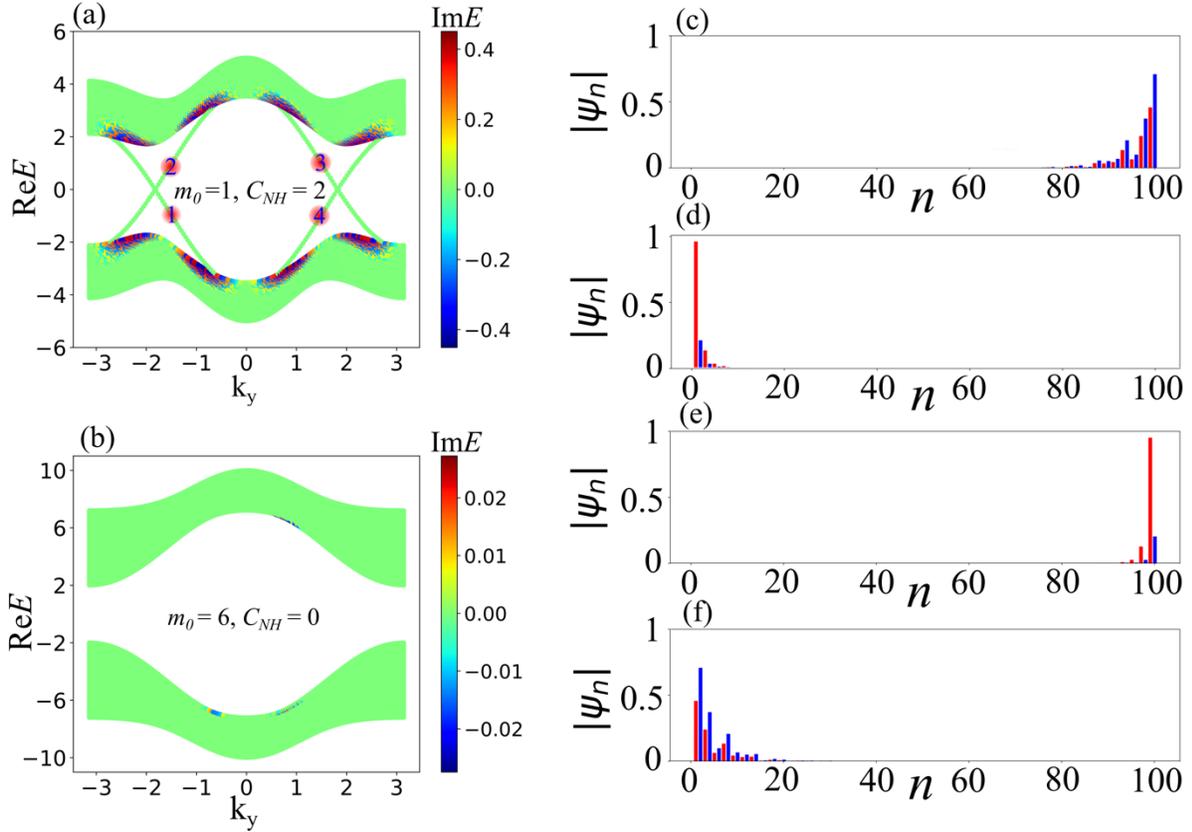

**FIG. 3.** (a-b) The complex energy spectrum for different values of $m_0$ for the supercell with periodic boundary conditions in the $y$-direction. (c-f) The wavefunction amplitude of the edge states indicated by the numbers 1 to 4 in (a). The number of sites along $x$-direction is $n_x = 50$, the amplitude of the wavefunction at state A (B) shown by the red (blue) histogram.

The Chern number in the Hermitian IQHI predicts number of unidirectional boundary states supported at an interface - the principle known as the bulk-boundary correspondence[42,43]. Similarly, the non-Hermitian Chern number is related to the net number of unidirectional edge states localized at the interface between topological and trivial system. To confirm the Bulk-edge correspondence in our non-Hermitian model, we study a system with periodic boundary condition in the $y$-direction and open boundary condition in the $x$-direction, thus forming a cylindrical (supercell) geometry. The energy spectrum for this geometry (as function of $k_y$) is shown in Fig. 3(a-b). For $m_0 = 1$, the system is in the topologically non-trivial phase with $C_{NH} = 2$, and there exist two pairs of edge states in the gap which, as shown in Fig. 3(c - f), are both localized to the edges of the cylinder. The spectrum of these states is purely real number, which implies that the edge modes of our non-Hermitian model are dissipationless. Inspection of the wavefunctions of the edge states reveals that for the states shown in Fig. 3(c,f) the phase difference between A and B components at each site is zero ($A$ and $B$ are in phase), while for the edge states in Fig. 3(d,e) the phase difference is $\pi$ ($A$ and $B$ are out of phase). As expected, no edge states are found in Fig. 3(b) corresponding to the topologically trivial phase, when non-Hermitian Chern number $C_{NH} = 0$. The band structure and edge states for cylinder geometry with periodic boundary in x-direction and open boundary condition in y direction are similar (See Supplemental Material Section II). In the Supplemental Material, we also calculate the non-Hermitian Chern number based on the open-bulk Chern number[44] with open boundaries in both x and y directions. The open-bulk Chern number's predictions are agree with the predictions calculated by our non-Hermitian Chern number Eq. (3). These observations confirm the Bulk-edge correspondence of our non-Hermitian model.

**Conclusions**

In this work, we proposed a two-band model of a non-Hermitian topological insulator supporting a higher valued non-Hermitian Chern number $C_{NH} = 2$, which, as the result, hosts two edge states with purely real spectrum in the topologically non-trivial band gap. More importantly, we develop a mathematical formalism and prove that the quantization of the non-Hermitian Chern number is protected by the generalized $C_4$ rotational symmetry present in our system and invoking the $\frac{\pi}{2}$ rotation along with the Hermitian conjugation operation (exchange of gain and loss). Our work shows the possibility to generalize the rotational symmetry, commonly used to describe Hermitian systems, and employ non-Hermitian character of systems to induce higher values of the topological invariants, such as the non-Hermitian Chern number. Our work thus outlines an approach to engineer physical systems supporting elevated number of edge states at the boundaries of non-Hermitian topological systems by judicious introduction of gain and loss, which may enable novel applications with robust guiding and multiplexing over multiple boundary states.

**Acknowledgements**
The work was supported by the NSF grants DMR-1809915 and OMA-1936351, the ONR award N00014-21-1-2092, and the Simons Collaboration on Extreme Wave Phenomena.

© kchen3@gradcenter.cuny.edu
∗ akhanikaev@ccny.cuny.edu

Kai Chen[1,2,3], Alexander B. Khanikaev[1,2,3]

[1]Department of Physics, City College of New York, New York, NY 10031, USA.
[2]Department of Electrical Engineering, Grove School of Engineering, City College of the City University of New York, 140th Street and Convent Avenue, New York, NY 10031, USA.
[3]Physics Program, Graduate Center of the City University of New York, New York, NY 10016, USA.

## I. Non-Hermitian Chern Number

In this section, we present the detailed derivations of non-Hermitian higher Chern number. For a Hermitian system with rotation symmetry can be represented by $\hat{H}(g\vec{k}) = \hat{U}_g \hat{H}(\vec{k}) \hat{U}_g^+$, $g$ is the generator of the corresponding rotation group, and $\hat{U}_g$ is the unitary representation of g. However, a non-Hermitian system, this relation could be violated since $\hat{H}(\vec{k}) \neq \hat{H}^+(\vec{k})$. In the main text, we consider a non-Hermitian tight-binding model with a generalized rotation symmetry described by $\hat{H}^+(g\vec{k}) = \hat{U}_g \hat{H}(\vec{k}) \hat{U}_g^+$, $g$ and $\hat{U}_g$ are the generator and corresponding unitary representation of rotation group $C_N$, i.e., $\hat{U}_g = e^{-i\frac{\pi}{n}m\frac{\hat{\sigma}_z}{2}}$, $m$ take integer values. In the main text, we take $C_4$ rotation group as a concrete example, therefore, $g$ is the $\frac{\pi}{2}$ rotation around z axis, i.e., $g = R_z\left(\frac{\pi}{2}\right)$, and in order to make our tight-binding model satisfy the generalized $C_4$ rotational symmetry we take $m = 2$. The non-Hermitian Berry connection $\vec{A}_n(\vec{k})$, Berry curvature $\vec{B}_n(\vec{k})$ and non-Hermitian Chern number $C_{NH}$ are defined as

$$\vec{A}_n(\vec{k}) = <\phi_n(\vec{k})| \nabla_{\vec{k}} |\psi_n(\vec{k})>, \tag{S.1}$$

$$\vec{B}_n(\vec{k}) = \nabla_{\vec{k}} \times \vec{A}_n(\vec{k}), \tag{S.2}$$

$$C_{NH} = \frac{1}{2\pi i} \iint \vec{B}_n(\vec{k}) \, d\vec{k}^2, \tag{S.3}$$

where $<\phi_n(\vec{k})|$ and $|\psi_n(\vec{k})>$ are the left and right eigenstates.

$\hat{H}^+(g\vec{k}) = \hat{U}_g \hat{H}(\vec{k}) \hat{U}_g^+$ imply that $\hat{U}_g^2 \hat{H}(\vec{k}) \hat{U}_g^{+2} = \hat{H}(g^2\vec{k}) = \hat{H}(-\vec{k})$, we have

$$\hat{H}^+(g\vec{k})\hat{U}_g|\psi_n(\vec{k})> = \hat{U}_g \hat{H}(\vec{k})|\psi_n(\vec{k})> = E_n(\vec{k})\hat{U}_g|\psi_n(\vec{k})>, \tag{S.4}$$

$$\hat{H}^+(g\vec{k})|\phi_n(g\vec{k})> = E_n^*(g\vec{k})|\phi_n(g\vec{k})>. \tag{S.5}$$

Using Eq. (S.4) and Eq. (S.5), we get $E_n^*(g\vec{k}) = E_n(\vec{k})$ and

$$|\phi_n(g\vec{k})> = e^{i\theta_g(\vec{k})} \hat{U}_g|\psi_n(\vec{k})>. \tag{S.6}$$

The left eigenstates at $g\vec{k}$ and right eigenstates at $\vec{k}$ are correlated by the unitary matrix $\hat{U}_g$. By virtue of the biorthogonal normalization condition of the left and right eigenstate of a non-Hermitian matrix, $<\phi_m(g\vec{k})|\psi_n(g\vec{p})> = <\phi_m(\vec{k})|\psi_n(\vec{p})> = \delta_{m,n}\delta(\vec{k}-\vec{p})$, we have

$$|\psi_n(g\vec{k})> = e^{i\,\theta_g(\vec{k})}\,\hat{U}_g|\phi_n(\vec{k})>, \quad (S.7)$$

and

$$|\psi_n(g^2\vec{k})> = e^{i\,\theta_g(\vec{k})}\,\hat{U}_g|\phi_n(g\vec{k})> = e^{i\,2\,\theta_g(\vec{k})}\,\hat{U}_g^2\,|\psi_n(\vec{k})>. \quad (S.8)$$

Then the group element $g$ map the non-Hermitian Berry connection to

$$\vec{A}_n(g\vec{k}) = <\phi_n(g\vec{k})|g^{-1}\nabla_{\vec{k}}|\psi_n(g\vec{k})> = <\psi_n(\vec{k})|U_g^+ e^{-i\theta_g(\vec{k})} g^{-1}\nabla_{\vec{k}}\,\hat{U}_g e^{i\theta_g(\vec{k})}|\phi_n(\vec{k})>$$

$$= g[-\vec{A}_n^*(\vec{k}) + i\,\nabla_{\vec{k}}\,\theta_g(\vec{k})], \quad (S.9)$$

$$\vec{A}_n(g^2\vec{k}) = <\phi_n(g^2\vec{k})|g^{-2}\nabla_{\vec{k}}|\psi_n(g^2\vec{k})> = <\phi_n(\vec{k})|\hat{U}_g^{+2} e^{-i2\,\theta_g(\vec{k})} g^{-2}\nabla_{\vec{k}}\,\hat{U}_g^2 e^{i2\theta_g(\vec{k})}|\psi_n(\vec{k})>$$

$$= g^2[\vec{A}_n(\vec{k}) + i\,2\,\nabla_{\vec{k}}\,\theta_g(\vec{k})], \quad (S.10)$$

Thus, the non-Hermitian Berry connection depends on the parity of the number of rotations, i.e.,

$$\vec{A}_n(g^l\vec{k}) = \begin{cases} g^l[\vec{A}_n(\vec{k}) + i\,l\,\nabla_{\vec{k}}\,\theta_g(\vec{k})] & for\ l \in 2\mathbb{Z} \\ g^l[-\vec{A}_n^*(\vec{k}) + i\,l\,\nabla_{\vec{k}}\,\theta_g(\vec{k})] & for\ l \in 2\mathbb{Z}+1 \end{cases}. \quad (S.11)$$

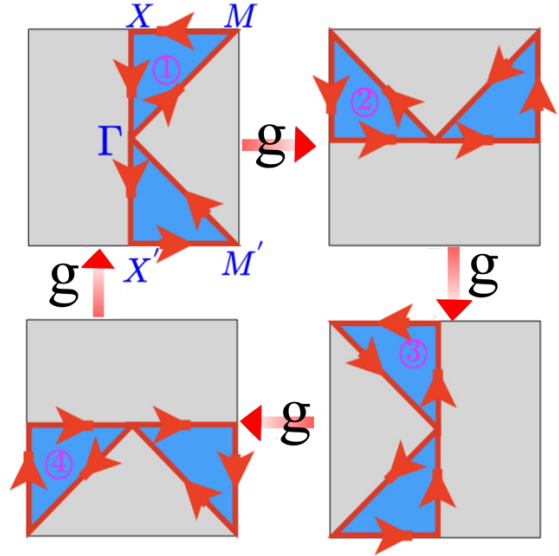

FIG. S1. The first Brillouin zone (BZ) and high symmetric points of the model.

The non-Hermitian Chern number

$$C_{NH} \equiv \frac{1}{2\pi i} \iint_{BZ} \vec{B}_n(\vec{k}) \, dk^2 = \frac{1}{2\pi i} \iint_1 \vec{B}_n(\vec{k}_1) dk_1^2 + \frac{1}{2\pi i} \iint_2 \vec{B}_n(\vec{k}_2) dk_2^2 + \frac{1}{2\pi i} \iint_3 \vec{B}_n(\vec{k}_3) dk_3^2 + \frac{1}{2\pi i} \iint_4 \vec{B}_n(\vec{k}_4) dk_4^2,$$

where {1,2,3,4} are the regions in the first Brillion zone as shown in Fig. S1.

$$C_{NH} = \frac{1}{2\pi i} \iint_1 \vec{B}_n(\vec{k}_1) dk_1^2 + \frac{1}{2\pi i} \iint_1 \vec{B}_n(g\vec{k}_1) \, g \, dk_1^2 + \frac{1}{2\pi i} \iint_1 \vec{B}_n(g^2 \vec{k}_1) \, g^2 \, dk_1^2 + \frac{1}{2\pi i} \iint_1 \vec{B}_n(g^3 \vec{k}_1) \, g^3 dk_1^2,$$

Using the definition of Berry connection $\vec{B}_n(\vec{k}) = \nabla_{\vec{k}} \times \vec{A}_n(\vec{k})$, and identity $\nabla_{\vec{k}} \times \nabla_{\vec{k}} \theta_g(\vec{k}) = 0$, we get

$$C_{NH} = \frac{1}{2\pi i} \iint_1 \vec{B}_n(\vec{k}_1) dk_1^2 - \frac{1}{2\pi i} \iint_1 \vec{B}_n^*(\vec{k}_1) \, dk_1^2 + \frac{1}{2\pi i} \iint_1 \vec{B}_n(\vec{k}_1) \, dk_1^2 - \frac{1}{2\pi i} \iint_1 \vec{B}_n^*(\vec{k}_1) \, dk_1^2$$
$$= \frac{2}{\pi} \, Im \iint_1 \vec{B}_n(\vec{k}_1) dk_1^2 = \frac{2}{\pi} \, Im \oint_\gamma \vec{A}_n(\vec{k}) d\vec{k}, \quad (S.12)$$

where the loop $\gamma \equiv (\Gamma \to M \to X \to \Gamma \to X' \to M' \to \Gamma)$. Combining Eq. (S.12) and relation $\int_{M \to X} \vec{A}_n(\vec{k}) d\vec{k} + \int_{X' \to M'} \vec{A}_n(\vec{k}) d\vec{k} = 0$, we get

$$C_{NH} = \frac{2}{\pi} \, Im\{\int_{\Gamma \to M} \vec{A}_n(\vec{k}) d\vec{k} - \int_{\Gamma \to M'} \vec{A}_n(\vec{k}) d\vec{k} - \int_{\Gamma \to X} \vec{A}_n(\vec{k}) d\vec{k} + \int_{\Gamma \to X'} \vec{A}_n(\vec{k}) d\vec{k}\}. \quad (S.13)$$

The value $\vec{k}$ in $\Gamma \to M'$ is correlated with the $\vec{k}$ in $\Gamma \to M$ by $g^2$, $g$ is the $\frac{\pi}{2}$ rotation in $\vec{k}$ space around $k_z$ axis.

$$\int_{\Gamma \to X'} \vec{A}_n(\vec{k}) d\vec{k} = \int_{\Gamma \to X} \vec{A}_n(g^2 \vec{k}) g^2 d\vec{k} = \int_{\Gamma \to X} \vec{A}_n(\vec{k}) d\vec{k} + i \, 2 \, [\theta_g(X) - \theta_g(\Gamma)], \quad (S.14)$$
$$\int_{\Gamma \to M} \vec{A}_n(\vec{k}) d\vec{k} = \int_{\Gamma \to M'} \vec{A}_n(g\vec{k}) g d\vec{k} = -\int_{\Gamma \to M'} \vec{A}_n^*(\vec{k}) d\vec{k} + i \, [\theta_g(M) - \theta_g(\Gamma)]. \quad (S.15)$$

Using Eq.(S.13), Eq.(S.14) and Eq.(S.15) we have

$$C_{NH} = \frac{2}{\pi} \, Im \, \{i \, 2 \, [\theta_g(X) - \theta_g(\Gamma)] + i \, [\theta_g(M) - \theta_g(\Gamma)] - \int_{\Gamma \to M'} (\vec{A}_n(\vec{k}) + \vec{A}_n^*(\vec{k})) d\vec{k}\}.$$

Finally, we get

$$C_{NH} = \frac{2}{\pi} [2\theta_g(X) - 3\theta_g(\Gamma) + \theta_g(M)]. \quad (S.16)$$

At high symmetry points $\vec{K}$ of Brillouin zone, $|\phi_n(g\vec{K})> = e^{i\theta_g(\vec{K})} \hat{U}_g |\psi_n(\vec{K})>$, and $1 = <\psi_n(\vec{K})|\phi_n(\vec{K})>$, we get $\theta_g(\vec{K}) = -Arg[<\psi_n(g\vec{K})|\hat{U}_g|\psi_n(\vec{K})>]$. For the case of $m = 2$, $\hat{U}_g = e^{-i\frac{\pi}{2}\hat{\sigma}_z}$, we get $\theta_g(\vec{K}) = \frac{\pi}{2} <\psi_n(g\vec{K})|\hat{\sigma}_z|\psi_n(\vec{K})>$. One should notice that $\theta_g(\vec{K})$ are expectation value of $\hat{\sigma}_z$ under right eigenstates.

## II. Solution of Cylinder Geometry

In the main article, we have discussed the cylinder geometry with periodic boundary condition in y-direction and open boundary condition in x-direction. Here, we study the edge states under the cylinder geometry with periodic boundary condition in x-direction and open boundary condition in y-direction. As shown in Fig. S2, the band structure and edge states of the two cylinder-geometries are similar.

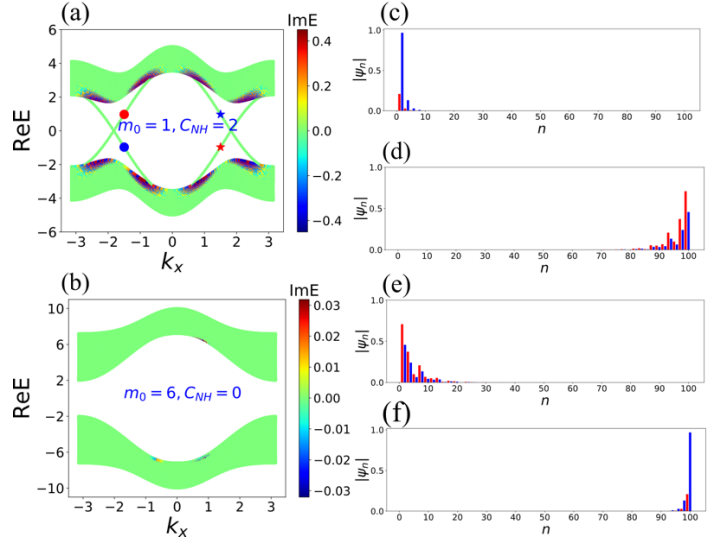

FIG. S2. (a-b) The complex energy spectrum for different values of $m_0$ for the supercell with periodic boundary conditions in the $x$-direction. (c-f) The wavefunction amplitude of the edge states indicated by the blue circle ((c)), red circle ((d)), blue star ((e)) and red star ((f)) in (a). The number of sites along $y$-direction is $n_y = 50$, the amplitude of the wavefunction at state A (B) shown by the red (blue) histogram.

### III. Solution of Open boundary Geometry and Real Space non-Hermitian Chern Number

In this section, we use the open-bulk Chern number (proposed in Ref. [1]) which motivated by the duality between the real space and Generalized Brillouin Zone (GBZ) to prove that our non-Hermition Chern number is a correct topological invariant for our non-Hermitian model with generalized rotational symmetry, although our non-Hermitian Chern number is defined by ordinary Brillouin Zone.

The open-Chern number of a band $\alpha$ is given as

$$C_\alpha = \frac{2\pi}{l_x l_y} tr(P_\alpha [[X_\alpha, P_\alpha], [Y_\alpha, P_\alpha]]). \quad (S.17)$$

where $P_\alpha$ is the bulk-band projection operator,

$$P_\alpha \equiv \sum_{n \in \alpha} |\psi_n\rangle\langle\phi_n|, \quad (S.18)$$

The open boundary system has a size $N_x \times N_y$, and $tr$ denotes the trace within the center region $l_x \times l_y$ with $l_j \equiv N_j - 2L_j$, $L_j$'s denote the thickness of boundary layers. The $P_\alpha$ take into the non-Hermitian nature of the problem which is highly sensitive to the boundary condition in the presence of non-Hermitian Skin effect[1].

We calculate the Eq. (S17) for our non-Hermitian model. The open-bulk Chern number of the lower band is shown in Fig. S3, which agree with the non-Hermitian Chern number calculated by our Eq. (S16). Therefore, our non-Hermitian Chern number is good topological invariant for our non-Hermitian model.

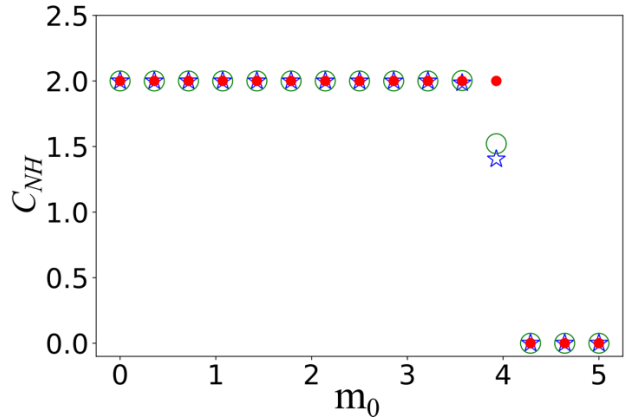

FIG. S3. Non-Hermitian Chern number for Eq.(S.17). Fixed parameters: $\gamma_z = 0.5$, and $t_x = t_y = t_z = 1$. The blue stars and the green circles are calculated by the open-bulk Chern number method, the size of square lattice is $N_x = N_y = 50$ (green circles) and $N_x = N_y = 30$ (blue stars); the truncation is $L_x = L_y = 10$. The red points are calculated by our non-Hermitian Chern number for Eq. (S.16).

## IV.  Open boundaries in both x and y directions

In this section, we investigate our non-Hermitian model with open boundaries in both x and y-directions. We numerically obtained complex spectra which are shown in Fig. S4. There exist edge modes in the open lattice for the topological gapped phase with non-Hermitian Chern number $C_{NH} = 2$.

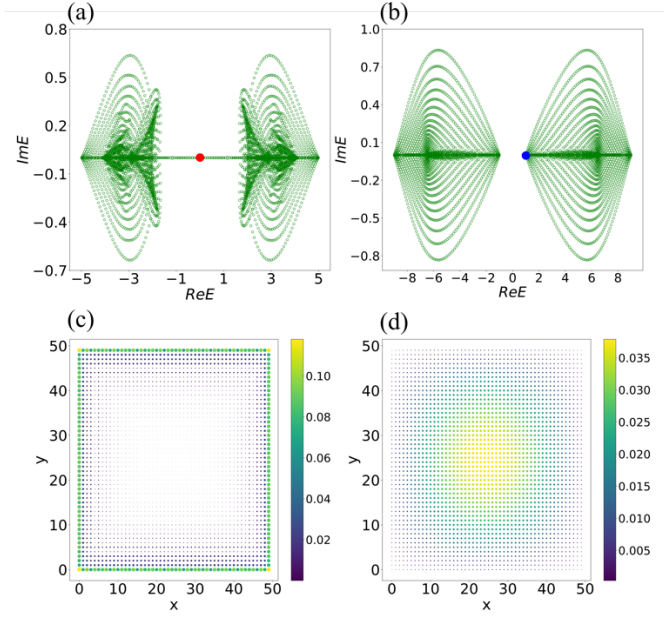

FIG. S4. Complex energy spectra and the profile of edge states of the non-Hermitian model given by the Eq. (2) in the main text which has open boundaries in both x and y directions. Fixed parameters: $\gamma_z = 0.5$, $t_x = t_y = t_z = 1$, and the size of the lattice is $N_x = N_y = 50$. Real and imaginary parts of the complex spectra are shown for (a) $m_0 = 1$ (corresponding to topological phase), (b) $m_0 = 5$ (corresponding to Trivial phase). (c,d) Amplitude distribution of eigenfunction corresponding the complex energy spectra indicated by the red point in (a) and blue point in (b).